\documentclass[runningheads]{llncs}
\usepackage[utf8]{inputenc}
\usepackage{graphicx}
\usepackage{hyperref, xcolor}

\usepackage{booktabs}

\begin{document}
\title{%
  Beginners' Quest to Formalize Mathematics:\\
  A Feasibility Study in Isabelle%
}
\titlerunning{Beginners' Quest to Formalize Mathematics}

\author{%
    Jonas Bayer\inst{2} \and %
    Marco David\inst{1} \and %
    Abhik Pal\inst{1} \and %
    Benedikt Stock\inst{1} %
}

\authorrunning{J. Bayer et al.}

\institute{
  Jacobs University Bremen, Campus Ring 1, 28759 Bremen, Germany.
  \email{\{m.david,ab.pal,b.stock\}@jacobs-university.de}
  \and
  Freie Universität Berlin, Institut für Mathematik, Arnimallee 3, 14195 Berlin, Germany.
  \email{jonas.bayer@fu-berlin.de}
}

\maketitle

\begin{abstract}
  How difficult are interactive theorem provers to use? We respond by
  reviewing the formalization of Hilbert's tenth problem in
  Isabelle/HOL carried out by an undergraduate research group at
  Jacobs University Bremen. We argue that, as demonstrated by our
  example, proof assistants are feasible for beginners to formalize
  mathematics. With the aim to make the field more accessible, we also
  survey hurdles that arise when learning an interactive theorem
  prover. Broadly, we advocate for an increased adoption of
  interactive theorem provers in mathematical research and curricula.

\end{abstract}

\section{Introduction}
\label{sec:intro}

The challenge to formalize all of mathematics, as issued by the QED Manifesto~\cite{qed-manifesto}, might
have seemed unrealistic for the 1990s but recent advances in theorem
proving clearly demonstrate the feasibility of using
theorem provers in mathematical research. Examples for this are the formalization of the odd-order theorem in
Coq~\cite{odd-order} and Kepler's conjecture in HOL
Light~\cite{flyspeck}. Even though these tools
provide the possibility of establishing mathematical truth once and
for all, mathematicians are reluctant to use interactive theorem
provers to verify the correctness of their proofs
\cite{benzmueller-editorial,royal-soc-preface}. ``Interactive theorem
provers are written by computer scientists for computer scientists,''
the complaint goes, quickly followed by a comment on their
infeasibility for non-experts.

In October 2017, twelve undergraduate students who just started their
university studies were asked to verify a mathematical proof using an
interactive theorem prover. Upon initiative of Yuri Matiyasevich,
whose contribution \cite{h10-book} was key to solving Hilbert's tenth
problem but who had no experience with proof assistants, the
undergraduates set out to formalize the problem and its solution.
Given the interactive theorem prover
Isabelle~\cite{isabelle02}\footnote{%
  In rest of the paper we write ``Isabelle'' to also mean 
  ``Isabelle/HOL''.} as ``relatively easy to learn,''
the \emph{Hilbert meets Isabelle} project was born.

Sixteen months and many ups and downs later, the project stands close
to completion. The students have made many mistakes and the large
workgroup has shrunk, but, most importantly, they all have learned a
lot. We herewith present a feasibility study of interactive theorem
provers for non-experts and disprove the concern raised earlier. From
young students to senior scientists in mathematics, computer science,
and engineering, everyone can pick up a proof assistant to formalize
their work --- it will be well worth the effort!

This paper reports about the ongoing project, reviews the tools and
resources that were used, and reflects on the learning process of the
group. With an emphasis placed on formalizing mathematics, we wish to
analyze the hurdles of becoming a proficient user of an interactive
theorem prover, scrutinize our mistakes, and share the lessons we
learned in the process. We also give a list of suggestions to
developers and future beginners to aid the interactive theorem proving
community to grow and welcome more mathematicians in the future.

\subsubsection{Overview} This paper is organized as follows: In
section~\ref{sec:background} we provide context to the formalization.
In particular, we briefly outline Hilbert's tenth problem and explain
the background and motivations of those involved. Then in
section~\ref{sec:mistakes} we analyze the process of formalization,
identify our key mistakes, the lessons learned from those mistakes,
and things we will do differently now. The current status of the
formalization is also given in this section. Finally, based on our
experience of learning Isabelle, in section~\ref{sec:recommendations}
we provide recommendations to the theorem proving community and
beginners interested in formalizing mathematics.

\section{The Quest to Formalize}
\label{sec:background}

On a visit to Jacobs University Bremen one and a half years ago, Yuri
Matiyasevich recruited students for a newly conceived research idea:
to conduct a formal verification of his solution to Hilbert's tenth
problem. In order to promote this project, he gave a series of talks
on the problem, its negative solution, and related questions
\cite{yumat17-tarski,yumat17-dirichlet}. These got a collection of
students curious and before long, a research group was formed. The
project was co-initiated by Dierk Schleicher who supported, mentored,
and supervised the workgroup.

However, neither Yuri Matiyasevich nor Dierk Schleicher had any
previous experience with interactive theorem provers. Coq~\cite{coq}
was known as a well established, yet difficult to learn proof
assistant, but Yuri Matiyasevich ultimately suggested Isabelle.
Supposedly with a less steep learning curve and better documentation,
this choice manifested. Thus began the quest to formalize.

\subsubsection{Hilbert's tenth problem and the MRDP theorem}
Hilbert's tenth problem comes from a list of 23 famous mathematical
problems posed by the German mathematician David Hilbert in
1900~\cite{hilbert-problems}. Hilbert's tenth problem asks about
Diophantine equations, which are polynomial equations with integer
coefficients: \emph{Does there exist an algorithm that determines if a
  given Diophantine equation has a solution in the integers?}
\cite{h10-book} The Matiyasevich--Robinson--Davis--Putnam theorem (also
known as the MRDP theorem, DPRM theorem, or Matiyasevich's theorem)
finished in 1970 by Yuri Matiyasevich~\cite{yumat70}, which states
that every recursively enumerable subset of the natural numbers is the
solution set to a Diophantine equation, implies a negative solution to
Hilbert's tenth problem.
 
For the proof, one first needs to develop the theory of Diophantine
equations. This entails showing that statements such as inequalities,
disjunctions, or conjunctions of polynomial equations can be
represented in terms of Diophantine polynomials. Then, as the first
major step in the proof, one shows that exponentiation also has such a
Diophantine representation. Next, after developing the notion of a
recursively enumerable set using a Turing-complete model of
computation (for instance using register machines), one shows that
this computation model, which accepts exactly the elements of
recursively enumerable sets, can be arithmetized, i.e. simulated using
Diophantine equations and exponentiation. Since there exist
recursively enumerable (semi-decidable), and hence Diophantine, sets
that are not decidable, any proposed algorithm would have to solve the
halting problem in order to decide an arbitrary Diophantine solution
set.

\subsubsection{Students' Background and Parallelization of Work}
\begin{figure}[h]
    \centering
    \includegraphics[width=\textwidth]{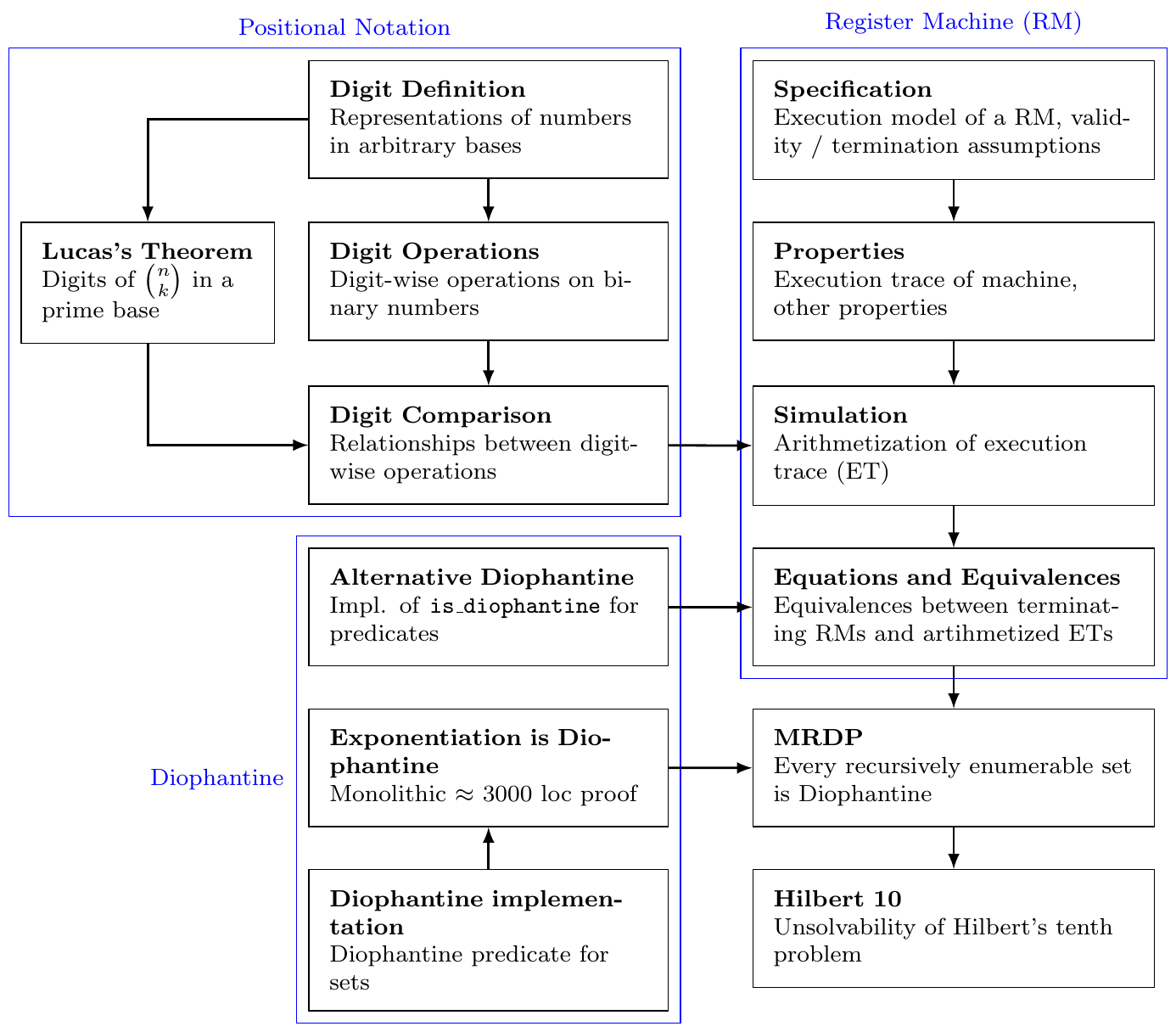}
    \caption{Simplified overview of the project's structure.}
    \label{fig:structure}
\end{figure}

After the team acquainted itself with the proof, the workgroup was
split accordingly: Team~I worked on showing that exponentiation is
Diophantine, Team~II on register machines and their arithmetization.
Figure \ref{fig:structure} gives an overview of the structure of the
project. For the first part, Matiyasevich \cite{h10-book} provides
detailed proofs; however, the arguments in the second part were at a
higher level of abstraction. While Team~I could work on formalizing
the first part with minimal Isabelle knowledge and the already
detailed paper proof, the second part of the formalization turned out
more challenging. The arithmetization of register machines required
not only an understanding of all details omitted in the paper, it also
required a good understanding of existing theories of already
formalized mathematics and practice with Isabelle's tools --- what
definitions lend themselves to automation? What is the appropriate
level of abstraction? What makes for a definition that can be used
well in proofs? etc.

Especially with the diverse background of many group members, the
above questions were not answered, let alone asked, immediately. The
students involved were mainly first year undergraduates studying
mathematics and computer science, who had not taken a course on
theorem proving. Not only did the students lack any foundational
knowledge in logic and type theory, some did not even have prior
programming experience. Combined, these factors resulted in an
approach to learning that can best be described as haphazard. However,
unbeknownst to the workgroup, these also became the preconditions for
a larger feasibility experiment in theorem proving --- how a group of
inexperienced undergraduates can learn an interactive theorem prover
to formalize a non-trivial mathematical result. In broader terms, the
next few sections report on this feasibility study.

\section{Sledgehammer Abuse, Foolish Definitions, and Reinventions}
\label{sec:mistakes}

In the beginning, many definitions, functions, lemmas, and proofs were
written without an overall understanding of their individual
functionality and utility. Especially for proofs, a lack of this
structural understanding prevented the formalization from advancing.
This section reviews the gamut of work done on the formalization
between October 2017 and February 2019 and analyses the key mistakes
that were made in the process, as well as the lessons learned from
them.

\subsubsection{Reinventions and Sledgehammer Abuse}
Due to the different nature of the two parts of the proof, the two
teams progressed with different speed and success. Team~I started by
implementing custom $2 \times 2$ matrices which are used frequently in the
first part. Although they initially searched for a matrix datatype
within existing Isabelle theories, this search turned out unsuccessful
as few relevant results appeared. And the results that were found did
not allow the team to infer how to use the respective implementations
for their own proofs. As the formalization progressed, many other
definitions and statements started relying on these custom types. Not
only did this result in a dependency-blow-up of elementary properties
that needed proving, they also prevented Isabelle from automating many
parts of the proofs.

The reimplementation of this basic type was followed by stating and
assuming intermediate lemmas without proof using Isabelle's convenient
\texttt{sorry} keyword. This allowed for parallelizing the work on
separate parts. The number-theoretic nature of proofs made it easier
to use tools like \texttt{sledgehammer} that call external automatic
theorem provers to search for proofs. The general approach to prove a
given statement was the following: state an intermediate step, then
check if \texttt{sledgehammer} can find a proof, otherwise introduce a
simpler sub-step and repeat.

Since the paper proof was understood in full detail and the internal
Isabelle libraries were sufficiently sophisticated, the
\texttt{sledgehammer}-and-repeat approach worked surprisingly well. In
fact, much of the entire first part was successfully formalized using
this approach. This, however, had two main flaws. First, the proofs
themselves were generated by automated theorem provers without human
insight, hence cumbersome and almost impossible to understand. Second,
since the approach worked relatively well it didn't incentivize the
members to learn more about the Isabelle system and understand the
functionality it provides. Remaining a mysterious tool that could
automagically prove theorems, \texttt{sledgehammer}'s capabilities,
limitations, and output were never actually understood.

\medskip

\subsubsection{Foolish Definitions}
In parallel to the above, Team~II worked on arithmetizing register
machines and the results of the second part of the proof, which
culminates in the statement that all recursively enumerable sets are
exponentially Diophantine. The groundwork underlying this part of the
implementation included a definition of register machines, in
particular Minsky machines. This modeling task initially posed a major
hurdle towards the formalization. In retrospect, the ideal
implementation makes all variables used in the proof become readily
accessible. The first implementation written was, however, the direct
opposite of that as it made extensive use of lists, fold operations,
and comprehensions. This approach, while easy to implement, turned out
to be too unwieldy for any proofs. In the end, the implementation from
Xu~et~al.~\cite{mech-turing} was used as model for the formalization.
While they describe a Turing machine, compatible ideas
were extracted and used to implement register machines.

Once a workable model of register machines was implemented, the group
could set on the goal to actually prove lemmas that were only stated
before. For Team~II, this is where the actual challenge of learning
Isabelle started all over. Although the register machine model
succeeded in being strongly modular, its properties were inherently
more complicated than the number-theoretic statements from the first
part. In particular, most lemmas about the workings of a register
machine typically required one to fix some initial state, some set of
instructions as well as to assume validity of all state transitions,
etc. Breaking proofs down into smaller and smaller pieces, as is
commonly done also in mathematics, hence became much more difficult.
In some sense, the large size of the implemented machinery posed a
mental barrier to tackling the stated lemmas. Extrapolating the
\texttt{sledgehammer}-and-repeat strategy of Team~I, Team~II initially
hoped for automated theorem provers to prove very extensive lemmas
without much human help. In retrospect this was a ridiculous
expectation.

\subsubsection{Expecting Intuition from Isabelle}
To add to this, it turned out that the small details of intermediate
proof steps were often not understood as well as the group thought.
This lead to ``proof-hacking'' scenarios even after lemmas had been
successfully split into smaller statements. Most prominently,
Matiyasevich~\cite[Section~4.4.2]{h10-book} gives a central property
of register machines without elaborate proof because it follows from
an analogous special case. Due to the similarity of the properties
both in writing and in function, this generalization was intuitively
clear. However, collectively the team did not know how to convey this
intuition to Isabelle. It took several months until a complete
``paper-proof'' of all intermediate steps, done by a member of the
group, could suddenly give the formalization of this property new
momentum. With a new straight-forward approach, its proof was
seamlessly completed. Even though many proofs are conceptual, often to
ease reading and understanding, every correct proof can be made formal
by definition, on paper and hence in an interactive theorem prover.

Finally, the exact implementation of finite power series used in the
proof posed one more difficulty. Throughout the project, three
definitions of such series coexisted. One can define a finite sum
directly, or alternatively define recursive functions which, in each
iteration, add a term to the series from the left or right. Their
equivalence can be easily proven; yet, depending on the specific use
case within a proof, the right definition becomes pivotal. Exactly the
above generalization benefited from an explicit definition of the
power series as a finite sum, and would have taken significantly more
effort with any of the other definitions. Incidentally, this is
similar to conventional mathematics on paper but contrasts
conventional programming where there often is no difference between
two equivalent definitions.

In similar fashion, the complexity of proofs may considerably change
depending on the facts which are added to the set of automatically
used simplification rules. As such, both the right setup of
definitions before proving as well as the right setup of the prover
determine the (human) provers' success.

\subsubsection{Current Status}
With all of this at hand, the second part of the formalization was
only recently advanced. Additionally, many conceptual issues were
resolved earlier this year with the dedication and input from Mathias
Fleury. As of writing, only few lemmas in the second part --- and
hence in the entire Isabelle formalization of the MRDP theorem ---
remain to be completed. In particular, these include more minor lemmas
on register machines, proving Lucas's theorem on digit-wise
representations of binomial coefficients in a prime base, and proving
that certain relations like binary digit-wise multiplication are
Diophantine. Table~\ref{tab:statistics} lists some statistics from the
current state of the formalization\footnote{The actual source code has
  been made available at \url{https://gitlab.com/hilbert-10/dprm}
  under the GPLv3 license.}. Once completed, the formalization is
expected to be sent as a submission to the Archive of Formal
Proofs~\cite{afp}.

\begin{table}[h]
  \centering
  \caption{Statistics about the current progress (as of commit
    \texttt{bea7403d}) of the formalization.}
  \label{tab:statistics}
  \begin{tabular}{l l}
    \toprule
    Lines of code                       & 7759 \\
    --- of which for Register Machines    & 2692 \\
    --- of which for Diophantine theories & 3856 \\
    --- of which for Positional Notation  & 1150 \\
    --- of which for miscellaneous files  & 61   \\
    \midrule
    Number of definitions          & 48 \\
    Number of functions            & 41 \\
    Number of lemmas and theorems  & 295 \\
    \bottomrule
  \end{tabular}
\end{table}

\subsubsection{Lessons Learned}

Throughout the above story, we\footnote{In rest of the text the
  authors use ``we'' to interchangeably refer to themselves as authors
  and as representatives of the workgroup.} learned many lessons which
we share below. From discussions with Isabelle users of different
background and at different levels, a small survey showed that these
also are issues for most learners. One could call the following
``trivial'' and we would probably agree. However, these lessons are so
essential that we recommend any future beginner to be absolutely aware
of them.

\begin{enumerate}
\item Merely understanding the idea of a proof and knowing how it is
  carried out conceptually does not suffice for its formalization. As
  tempting as is might seem to start proving in Isabelle, the
  formalization should only be started after the proof has carefully
  been written down on paper in full detail.

\item Working with concepts that frequently pop up in mathematics, it
  is likely that someone else has worked on them before. Instead of
  reinventing the wheel, one should search the existing and extensive
  Isabelle libraries.
  
\item The exact implementations of functions and predicates can both
  facilitate but also impair the progress of any proof. The chosen
  definitions directly reflect the approach taken to the problem,
  which also has a big impact on conventional proofs. However, they
  additionally require an adequate level of abstraction so that human
  and proof assistant can work with them effectively.
\end{enumerate}

\subsubsection{What to Change Next Time}

Conjointly, reflection of our method of working and learning reveals
several defects, which are presented below. We suspect that these, in
turn, are very likely to have systematically caused the above
mistakes, or at least delayed their mending. In particular, we view
the following points as definitive ``not~to-dos'' for any
formalization project using interactive theorem prover.

The most valuable source for beginners is undoubtedly Tobias Nipkow's
Concrete Semantics \cite{concrete-semantics}. We would have learned
much quicker and with more structure, had we had strictly followed
this book and its exercises. Learning Isabelle on the fly, in a
\emph{learning-by-doing} fashion, and looking up commands as needed
was futile and it remains questionable if any such method of learning
can be successful. Given many group member's previous programming
experience, we clearly overestimated our ability to transfer this
knowledge to an interactive theorem prover.

Expanding on this note, our experience suggests that relying on
programming experience is helpful but should be done in tandem with an
awareness of the key differences between programming and proving. Most
notably, theory files are not compiled nor executed and proofs need to
be written with a much more mathematical and structured mindset, as
compared to programming. Interactive theorem provers are not just
``yet another programming language'' and our failure to realize this
has only lengthened the learning process.

We became aware of the two mistakes described above after connecting
with the very approachable Isabelle community. Only then we realized
how na\"{i}ve our initial approach to learning Isabelle was. Hence,
when working on a formalization for the first time, it is very helpful
to have an expert around who can be consulted when more conceptual
questions arise. We agree that this may not be ideal, which is why
further discussion on this issue follows in the next section.

\section{Decoupling Learning from Experienced Individuals}
\label{sec:recommendations}

We all have learned a lot from different members of the Isabelle
community. In this ongoing process, we gradually realize that
there's more to interactive theorem proving than just having one's
lemmas accepted by the computer. A prime example for this is knowing
what definitions are useful in which scenario. We've observed
both ourselves and others, that experienced users seem fundamental to
one's Isabelle education. Most of this education goes beyond mere
factual information and includes understanding the Isabelle system on
a deeper level, developing a systematic methodology of writing proofs,
and developing a ``feeling'' for the proof assistant. From this, we
conjecture the following.

\begin{quote}
  \emph{Conjecture.} Learning Isabelle currently depends on having an
  experienced user in reach who can regularly answer questions.
\end{quote}

\noindent%
We speculate that this can be generalized to other interactive theorem
provers, too. While we agree that learning from another user or
developer in person is certainly efficient, this becomes unsustainable
as the community grows. This naturally begs the question: \emph{How
  can a beginner's learning process become more guided by resources
  and documentation, therein more independent?} We do not, and
possibly can not, answer this question exhaustively. Nevertheless, we
ask this question both to ourselves and the community and present our
attempt at answering it.

\subsubsection{Expand Documentation}
Documentation plays a key role in helping new users get accustomed to
a new tool. And accessible, readable and easy to navigate
documentation, hence, is key to promote self-learning. As beginners
our first point of contact with Isabelle's documentation system was
the ``Documentation'' tab in the prover IDE. However, we found it
difficult to navigate as there was no clear indication of which
document is suitable for beginners. In retrospect we realize that
working though Nipkow's tutorial \cite{concrete-semantics} would have
been the most ideal. However, we still feel that the current
documentation system could be expanded as follows.

We identify four key parts of a systematic documentation
system\footnote{This identification is, by no means, original. Many
  large open-source software projects are aware of this structure and
  routinely advocate for documentation that conforms to it. See for
  instance the Django documentation \cite{django} and the Write the
  Docs project~\cite{wtd}.}. Tutorials that walk new users through
specific parts of the Isabelle system, how-to guides for learning and
using tools, topic guides that give the theoretical basis for many of
the features, and finally a repository of references that document all
necessary details. While the current documentation system addresses
three of those parts, it still lacks a crucial link that connects
them: the topic guide. This lack immediately implies that any deeper
understanding of the system can only come from being around regular
users -- hence tightly coupling a successful learning experience to
advanced users.

\subsubsection{Maintain a Knowledge-Base}
In tandem with documentation, it helps to accumulate a knowledge-base
of beginner- and intermediate-level questions and answers. The
Isabelle-users mailing list currently hosts the entire range of
questions and is definitely appropriate for advanced questions.
However, the thread archives are not suitable nor effectively
searchable as a database for questions which many more are likely to
have. Hence, we encourage users to ask these questions on Stack
Overflow or a similar online forum. Stack Overflow, for example,
aspires to become an ultimate and exhaustive knowledge base, and has
achieved this for many larger communities. Conventional programming
languages strongly benefit from that any basic question --- as
elementary it may be --- has been asked and answered before. The
interactive theorem proving community can do this as well.

We suggest that introductory-level resources like Concrete Semantics
or the Isabelle Community Wiki main page actively encourage users to
ask questions on Stack Overflow to build such a knowledge-base. This
way, every question will only need to be asked and answered once, and
everyone can benefit from users who share their expertise.

\subsubsection{Develop the Isabelle Community Wiki}
Thirdly, we suggest to expand the Isabelle Community Wiki in a similar
fashion: by an organic community effort. In the initial stages of our
project, we used an internal ``Isabelle Cheat Sheet'' to facilitate
mutual knowledge exchange. This Cheat Sheet was meant to be a platform
where common problems and their solutions could be presented to
everyone. In this respect, the intention was very similar to the
``Isabelle Community Wiki''. Although, interestingly enough, our own
Cheat Sheet and the Wiki were completely disjoint until we merged
them.

While the Cheat Sheet initially only included very basic syntactical
facts it was quickly extended by features of Isabelle that are not
described in existing beginner-level resources, e.g. Concrete
Semantics. This includes the possibility of passing arguments to
\texttt{sledgehammer} or how to look up facts in the existing
theories. Other facts on the Cheat Sheet include keywords that can be
used for custom case distinctions. Coming back to the previous point, we found
out about the latter specifically after asking the Isabelle community
on Stack Overflow. The fact that the question and answer have received
several upvotes indicates that questions like this are indeed relevant
to a broader audience.

\medskip 

\subsubsection{Adopt into University Curricula}
In a larger scope, all of the aforementioned would strongly benefit
from a growing user base. Having a larger community means that more
people will ask questions and thereby create documentation, as well as
eventually become experts themselves, working on exciting projects.
As a matter of fact, knowing how to use an interactive
theorem prover can be valued highly in many fields. Clearly, there is
academia with mathematics and computer science which both have an
interest and sometimes even a need to formally verify \cite{flyspeck}.
But uses in industry and engineering are equally compelling: formally
verified robots, airplanes, rockets, and nuclear plants prove
attractive to many companies and governments. Just one example of this
prevailing relevance is given by the annual NASA Formal Methods
Symposium.

In order to connect potential new users to the interactive theorem
proving community as early as possible, we think that initiatives like
\emph{Proving for Fun} \cite{proving-fun}, i.e. ``competitive
proving'' challenges, are a great idea to popularize the m\'{e}tier.
Well-established competitive programming contests range from the
International Informatics Olympiad to tech giant sponsored events and
attract students as young as middle school from all over the world.

More radically, we suggest this subject be adapted into mathematics,
computer science, and engineering curricula at universities. For wide
acceptance, in particular Bachelor students need to be exposed to
these tools before they specialize. Otherwise, knowledge will keep
being passed on from PhD student to PhD student within existing
research groups, but not become decoupled from exactly these. Of
course, this integration can happen step by step. Initially, there may
be an small elective course on interactive theorem proving, or some
part of a current course on logic can be dedicated to introducing an
interactive theorem prover. Once this exists, much more becomes
possible. In mathematics classes, there can be extra-credit problems
for also formally verifying one's proof, or eventually a full exercise
with this purpose. Theorem proving helps to teach what theorems and
facts are precisely used in every step of a proof \cite{buzzard}.
 
Some might classify this as a significant change in paradigm for
university-level education. We argue that our suggestion may well be
compared to computer algebra systems, which just entered Bachelor
curricula less than two decades ago. In this regard, interactive
theorem provers are the logical next step. With Mathematica, SAGE, and
similar systems successfully assisting computation and visualization,
it's now time to introduce interactive theorem provers like Isabelle
to assist modelling and proving. Initially well-suited as educational
tools, they might eventually also make their way into day-to-day
research work.

\section{Conclusion}
\label{sec:conclusion}

Our experience shows that non-experts can indeed learn interactive
theorem proving to an extent that allows them to formalize significant
mathematical results. Within one and a half years, we gained enough
Isabelle proficiency to formalize a core part of the solution to
Hilbert's tenth problem. We are happy to have used Isabelle for this
purpose, which we found to have a modest learning curve and be worth
the time investment. To future projects of similar kind, we recommend
that beginners approach learning an interactive theorem prover in a
more structured way than we did. To this end, we found Tobias Nipkow's
``Concrete Semantics'' \cite{concrete-semantics} the most helpful
first introduction to Isabelle. In general, we recommend to use a
single beginner-friendly resource which should also be clearly
advertised as such by more experienced members of the interactive
theorem proving community. For carrying out a formalization, we
realize that it is most crucial to start with a detailed ``paper
proof'' in order to then verify every single step in a proof
assistant.

Moreover, we find that interactive theorem provers are attractive to
many more fields and industries than their current user-base. Notably,
the group we encourage most to adopt proof assistants are
mathematicians, not least by incorporating them into university
curricula. Our feasibility study showed that interactive theorem
proving is doable and practical --- now it is the time to start
formalizing mathematics on a larger scale.

\bigskip

\subsubsection*{Acknowledgements}

We want to thank the entire workgroup~\cite{h10-floc18}, without whose
involvement we wouldn't be writing this paper; as well as Yuri Matiyasevich for initiating and guiding the project. Moreover, we would like
to express our sincere gratitude to the entire welcoming and
supportive Isabelle community. In particular we are indebted to
Mathias Fleury for all his help with Isabelle. Thank you also to
Christoph Benzmüller for mentoring us as well as Florian Rabe for
suggesting this contribution and helping us prepare the final version.
Furthermore, we thank everyone who replied to our small survey,
sharing their experience and opinion on this topic with us. Finally, a
big thank you to our supervisor Dierk Schleicher, for motivating us
throughout the project, connecting us to many experts in the field,
and all his comments on this article.

\bibliographystyle{splncs04}
\bibliography{ref}
\end{document}